# Electromagnetic Spatiotemporal Differentiators

**Yi Zhou[1], Junjie Zhan[1], Ziyang Xu[1], Yifan Shao[1], Yubo Wang[1], Yongdi Dang[1], Sen Zhang[1], Yungui Ma[1,*]**

[1]State Key Lab of Modern Optical Instrumentation, Centre for Optical and Electromagnetic Research, College of Optical Science and Engineering; International Research Center for Advanced Photonics (Haining Campus), Zhejiang University, Hangzhou 310058, China

*Corresponding author: yungui@zju.edu.cn

**Spatiotemporal optical computing devices which could perform mathematical operations in both spatial and temporal domains can provide unprecedented measures to build efficient and real-time information processing systems. It is particularly important to realize the comprehensive functions in a compact design for better integration with electronic components. In this work, we experimentally demonstrated an analogue spatiotemporal differentiator in microwaves based on an asymmetrical metasurface which has a phase singularity in the spatiotemporal domain. We showed that this structure could give rise to a spatiotemporal transfer function required by an ideal first-order differentiator in both spatial and temporal domains by tailoring the unidirectional excitation of spoof surface plasmon polaritons (SSPPs). The spatial edge detection was performed utilizing a metallic slit and the temporal differentiation capability of the device was examined by Gaussian-like temporal pulses of different width. We further confirmed the differentiator demonstrated here could detect sharp changes of spatiotemporal pulses even with intricate profiles and theoretically estimated the resolution limits of the spatial and temporal edge detection. We also show that the pulse input after passing the spatiotemporal differentiator implemented here could carry a transverse orbital angular momentum (OAM) with a fractal topology charge which further increases the information quantity.**

The rapid development of intelligent informatic technologies is revolutionizing our daily life based on mega data computations mostly carried out in electronical ways. However, the correlated issues with respect to operation speed and energy consumption will be more urgent when the scenarios are broadened to transiently process ever-increasing databank. In the past years, as an alternative, analogue optic computing have evolved into a newly appealing research field by virtue of its extraordinary advantages including low-energy consumption, highly integrable capability, and natural parallel processing characterisitic.[1] Harnessing artificial electromagnetic materials sophisticatedly engineered to achieve peculiar mathematical calculations without analogue-to-digital and digital-to-analogue converters, analogue optical computing devices can overcome inherent drawbacks of modern digital computing systems in terms of heat generation, operation speed, energy consumption.[1–5] Moreover, analogue optical computing devices outperform traditional analogue computing systems leveraging mechanical or electronic components with respect to structure complexity and response speed.[6]

According to the operational domain, analogue optical computing is generally categorized into two kinds, namely analogue optical spatial computing and analogue optical temporal computing.

Analogue optical spatial computing operating on electromagnetic fields with respect to their spatial profile can be realized by a 4-F system or Green's function method.[6] For the former, the electromagnetic field is transformed by the spatial Fourier transform lenses and processed with a spatial frequency filter in the spatial frequency domain. The separate design gives the necessary degrees of freedom to optimize the performance of the whole system, which has been vastly explored to demonstrate numerous optical computing devices.[6–13] Comparatively, Green's function method can be manipulated to build lighter and more compact optical computing systems by directly implementing mathematical operations in the real space without additional elements. Leveraging the nonlocal response of artificial electromagnetic materials to achieve desired spatial transfer function of a specific mathematical calculation, Green's function method have been employed to establish versatile analogue optical spatial computing devices.[6,14–39] Among these devices, analogue optical spatial differentiators[6,14–32] have attracted increasing attention due to their application potential in the fields of edge detection and image processing. Analogue optical temporal computing which handles the pulse envelope with respect to time has been studied for several decades. Researchers have developed various related devices by engineering their temporal transfer functions based on wavelength-scale structures.[40–48]

Recently, to combine the spatial and temporal field operation functions into a single device is attracting more interests, i.e., realizing analogue optical spatiotemporal computing[49–60]. This evolution enables the optical hardware to process the field signals in both spatial and temporal domains, substantially enhancing the parallel operation capability and channel bandwidth. For example, dielectric grating or multilayer structure was ever proposed to fulfill the optical spatiotemporal differential operations in different orders by Golovastikov et al.[49] and the authors[53], respectively. Inspired by the recent observation of transverse orbital angular momentum (OAM) in the spatiotemporal domain[61], the first-order spatial and temporal differentiation has gained special attentions mainly employing asymmetrical grating-type structures made of metals[56] or dielectrics[58] which can generate a zero transfer function with a $2\pi$ winding phase in both Fourier spaces of the spatiotemporal domains. Compared with high-order differentiation, in this case, for input spatiotemporal pulse, the output pulse not only gives the spatial and temporal profiles of the input but also carries a transverse OAM, enhancing the information quantity[58]. But so far, no experiment has ever been reported to prove this prominent function.

In this work, we experimentally demonstrated an analogue spatiotemporal differentiator in microwaves utilizing asymmetric grating-type metasurface. As discussed before, this type of mirror-breaking photonic structures can produce a phase singularity in the spatiotemporal domain under unidirectional excitation of spoof surface plasmon polaritons (SSPPs)[56] or Fano line-shape resonance[57]. The spatiotemporal transfer function for our specifically optimized device resembles the kernel of an ideal first-order spatiotemporal differentiator. All these features and the relevant computing functionalities are systematically inspected in both spatial and temporal domains theoretically and experimentally. In the end, we also point out that the transmitted pulse field of our device could also carry a transverse OAM with a fractal topology charge thus further enhancing the information quantity.

**Results**

**Design of electromagnetic spatiotemporal differentiator.** Fig. 1 schematically illustrates the analogue electromagnetic spatiotemporal differentiator based on a bilayer metal grating. Considering an incident pulse $a_{\text{in}}(x, t)$ with the TM-polarization (magnetic field along the $y$ axis), the transmitted pulse $a_{\text{out}}(x,t)$ could be attained by Fourier analysis method with the transfer function of the metasurface device. The incident (transmitted) spatiotemporal pulse would be decomposed into a series of monochromatic plane waves employing the spatiotemporal Fourier transform as[56,58]:

$$a_{\text{in(out)}}(x,t) = p_{\text{in(out)}}(x,t)\exp(-i\omega_0 t)$$
$$= \iint \tilde{p}_{\text{in(out)}}(k_x,\Omega)\exp(ik_x x - i\Omega t)\mathrm{d}k_x\mathrm{d}\Omega \qquad (1)$$

where $p_{\text{in(out)}}(x,t)$ and $\tilde{p}_{\text{in(out)}}(k_x,\Omega)$ denote the pulse envelope and the corresponding envelope spectrum of the incident (transmitted) wave, respectively. $\omega_0$ denotes the center angular frequency of the spatiotemporal pulse. $k_x$ denotes the in-plane wavevector along the $x$ direction. $\Omega$ denotes the sideband angular frequency, i.e., $\Omega = \omega - \omega_0$. The transfer function (here transmission coefficient) of the device could be expressed as:

$$H(k_x,\Omega) = \tilde{p}_{\text{out}}(k_x,\Omega)/\tilde{p}_{\text{in}}(k_x,\Omega) \qquad (2)$$

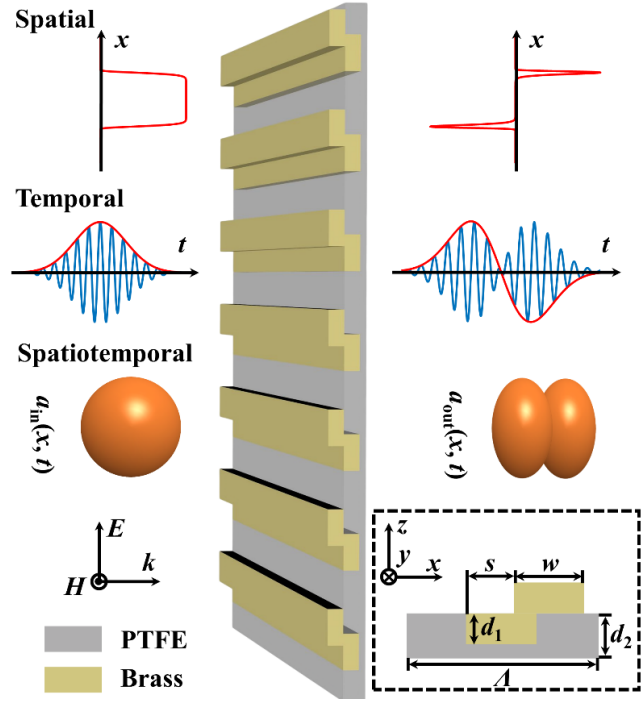

**Fig. 1 Schematic diagram of the analogue electromagnetic spatiotemporal differentiator utilizing a bilayer metal grating.** For TM-polarized incident pulse $p_{\text{in}}(x,t)$, the differentiator could perform first-order spatiotemporal differentiation. The cross-section view of unit cell of the metal grating composed by brass strips and PTFE substrate is shown in the inset.

The cross-section view of the unit cell of the metasurface is shown in the inset of Fig. 1. The metasurface is composed by two metal sub-gratings which have the same width $w$ and thickness $d_1$. We chose brass as the metal material due to its high conductivity in the microwave regime. Polytetrafluoroethylene (PTFE) substrate with the thickness $d_2$ was adopted because of its tractability and nearly lossless in the frequency band of interest. The transfer function of the first-order spatial differentiation ($H \propto ik_x$) has odd phase response with respect to $k_x$, which requires that the differentiator should break the mirror symmetry with respect to *x* and *z* axes simultaneously[19]. Such requirement would be realized by adjusting the lateral displacement *s* between two metal sub-gratings and embedding one sub-grating in the PTFE substrate. There is only zero-order diffraction for the normal incidence when the period of the unit cell *Λ* is smaller than the center wavelength in vacuum.

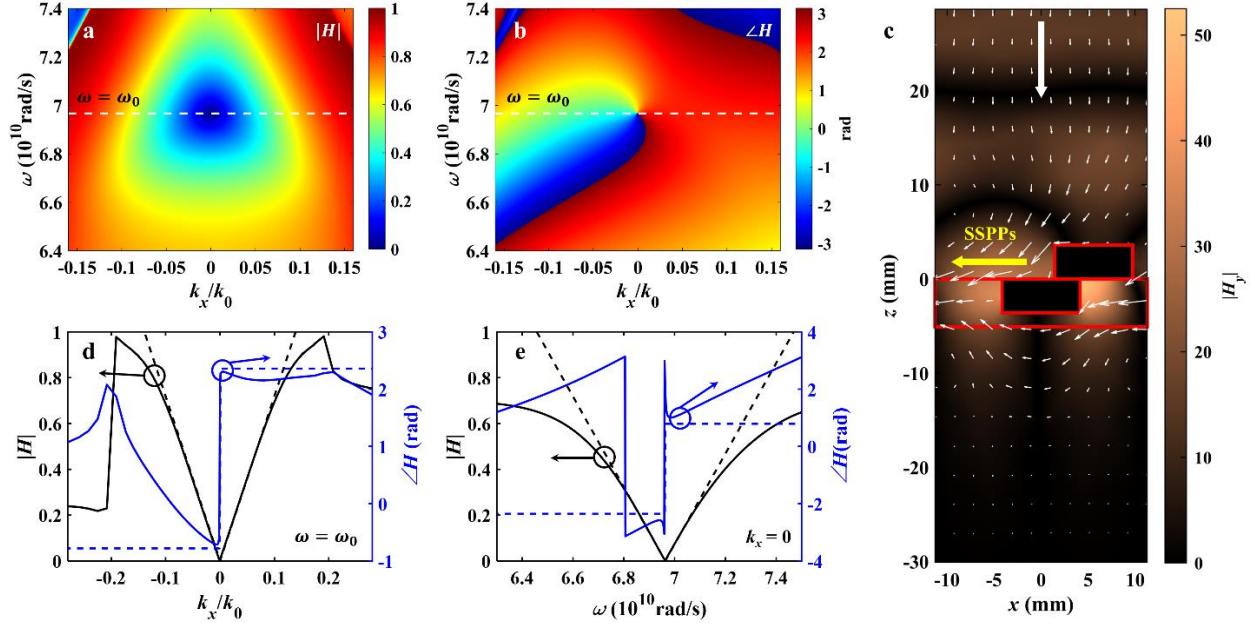

**Fig. 2 Simulation results of the first-order analogue spatiotemporal differentiator. a** Amplitude |*H*| and **b** phase ∠*H* distributions of the transfer function with respect to the normalized in-plane wavevector $k_x/k_0$ and angular frequency *ω*. $k_0$ is the wavevector in vacuum. $\omega_0$ is the angular frequency of the resonant mode for normal incidence. **c** Time-averaged Poynting vector (small white arrows) and amplitude distribution of the magnetic field $|H_y|$ (colormap) with $\omega = \omega_0$ for normal incidence. The large white arrow and yellow arrow show the propagation directions of the incident wave and SSPPs, respectively. The red lines show the outline of the unit cell of the metasurface. **d-e** Amplitudes (black) and phases (blue) of the transfer function along $\omega = \omega_0$ (white lines in **a-b**) and $k_x = 0$. The solid and dashed lines correspond to the transfer functions of simulation results and the fitting ones with Eq. 3, respectively. The black and blue arrows represent the curves to the left and right *y*-axis.

The geometry parameters of the metal grating were optimized as *Λ* = 22.6 mm, *w* = 8.3 mm, *s* = 5.6 mm, $d_1$ = 3.6 mm and $d_2$ = 5.08 mm. The Frequency Domain Solver of the commercial full-wave simulation software CST Studio Suite was employed to simulate the electromagnetic response of the device. The electric conductivity of brass was set to $3 \times 10^7$ S/m. The dielectric constant and loss tangent of PTFE were set to 2.2 and 0.001, respectively. We adopted Unit Cell boundary conditions along the *x*-axis and *y*-axis and Open boundary condition along the *z*-axis. Amplitude and phase distributions of the transfer function *H* with respect to the normalized in-plane wavevector $k_x/k_0$ and angular frequency *ω* are given in Figs. 2a-b, respectively. Here $k_0$ is the wavevector in vacuum. A significant resonant mode is excited with $\omega_0 = 6.96 \times 10^{10}$ rad/s for normal incidence. Previous researches have proved that metasurfaces could be leveraged to realize the unidirectional excitation of SSPPs with the phase matching condition satisfied[62]. Fig. 2c provides the time-averaged Poynting vectors and the amplitude distribution of the magnetic field $|H_y|$ with $\omega = \omega_0$ for normal incidence. The pronounced confinement feature of SSPPs is observed near the metal grating. The unidirectional excitation of SSPPs along the $x$-axis direction at the metal-air interface is validated by scrutinizing the time-averaged Poynting vectors. According to the coupled-mode theory, the transmitted field of the metasurface is the coupling between the direct transmission and the leakage radiation of SSPPs[16,56]. The electromagnetic responses of SSPPs are similar to those of SPPs. The transfer function around the spoof surface plasmon resonance could be approximated as[56]:

$$H(k_x, \Omega) \approx c_x k_x + c_t \Omega \tag{3}$$

where $c_x$ and $c_t$ are two different complex numbers. When there is a phase difference between $c_x$ and $c_t$, the phase singularity will appear in the phase distribution of the transfer function[58]. The transfer function has the linear form both for $k_x$ and $\Omega$, which tallies with requirements of the first-order spatiotemporal differentiation. Therefore, the transmitted pulse envelope can be calculated by the following equation:

$$p_{\text{out}}(x,t) = -\mathrm{i}c_x \frac{\partial p_{\text{in}}(x,t)}{\partial x} + \mathrm{i}c_t \frac{\partial p_{\text{in}}(x,t)}{\partial t} \tag{4}$$

Thus, the spatiotemporal differentiator can realize the first-order differentiation for the incident pulse envelope in both spatial and temporal domains.

The amplitudes and phases of the simulated transfer function of the metal grating as a function of in-plane wavevector and angular frequency at $\omega = \omega_0$ and $k_x = 0$ are given by black and blue solid lines in Fig. 2d-e, respectively. When $-0.09 \leq k_x/k_0 \leq 0.09$ and $-0.025 \leq \Omega/\omega_0 \leq 0.015$, the amplitudes of the transfer function exhibit a nearly linear relationship with both $|k_x|$ and $|\Omega|$. Moreover, the phases of the transfer function undergo a ~$\pi$ difference sharp change near the critical conditions of $k_x = 0$ and $\omega = \omega_0$. These characteristics coincide with the demands of the first-order spatial and temporal differentiation. By setting $c_x = 7.20 \exp(2.36\mathrm{i})/k_0$ and $c_t = 13.77 \exp(0.79\mathrm{i})/\omega_0$, we find that the transfer function of the metasurface could be well fitted utilizing Eq. 3, as indicated by the dashed lines in Fig. 2d-e. Considering $c_x/c_t \approx 0.52c \times \exp(\pi\mathrm{i}/2)$ ($c$ denotes the light velocity in vacuum), the transfer function of the structure exhibits the phase singularity with a $2\pi$ phase winding around it in the $k_x$~$\omega$ plane shown as Fig. 2b. In an enclosed loop in the vicinity of $k_x = 0$ and $\omega = \omega_0$, this phase singularity leads to zero amplitude for the transfer function[58].

**Experimental realization.** The schematic diagram of the experimental setup for measurements of the transmission coefficient (namely transfer function) of the metasurface sample is given in Fig. 3a. The vector network analyzer VNA (ZVA 40, Rohde and Schwarz) connected with a pair of antennae (SHA120S20, MWT) via phase stable cables. In the anechoic chamber, the receiving antenna A2 captured microwave signals radiated from the transmitting antenna A1. The fabricated metasurface sample, which was displaced from A1/A2 approximately by a 50 cm distance away, was mounted on a manual rotation stage RS (RBB300A/M, Thorlabs). In the experiment, the precision of the rotation stage was controlled at ~0.5°. The transmitted signal with phase compensation without the sample was considered as the reference of the transmission coefficient. Fig. 3b presents the photograph of the fabricated metasurface. We separately manufactured periodic grooves on PTFE substrate and brass strips using the computer numerical control (CNC) method. Then, the sample was fabricated by assembling the PTFE substrate and brass strips with screws. The size of the fabricated sample is 150×600 mm$^2$.

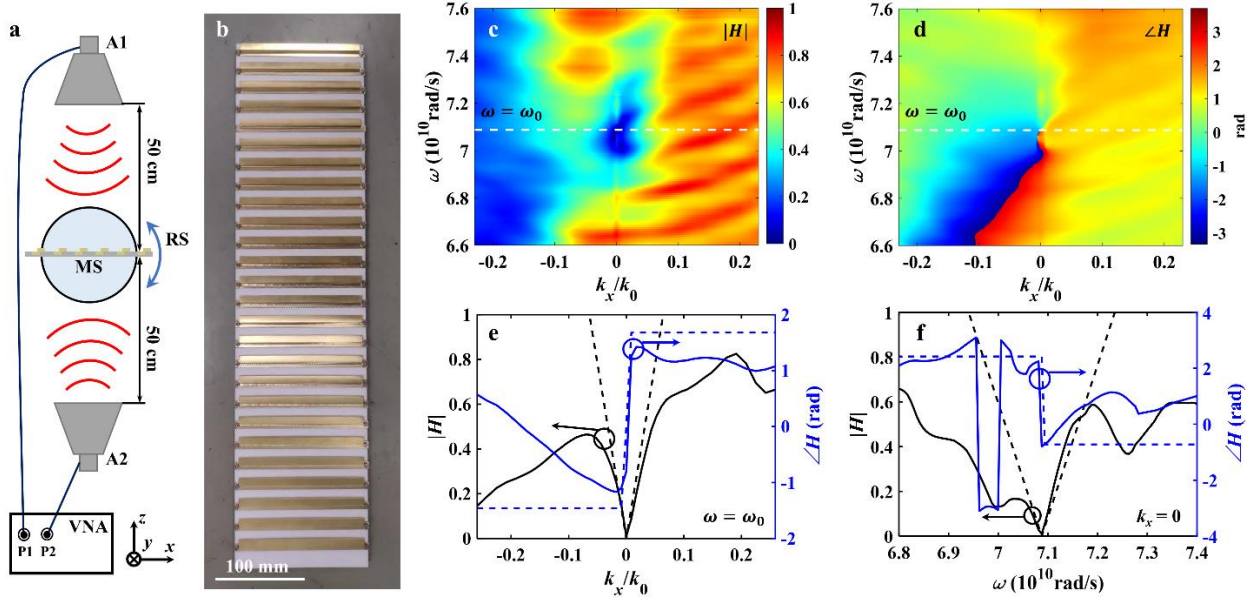

**Fig. 3 Experimental measurements of the transfer function of the fabricated metasurface sample. a** Schematic diagram of experimental setup utilized to measure the transfer function of the metasurface. A1/A2: transmitting/receiving antenna, MS: metasurface, RS: manual rotation stage, VNA: vector network analyzer, P1/P2: ports of VNA. **b** The photograph of the fabricated metasurface sample. **c** Amplitude $|H|$ and **d** phase $\angle H$ distributions of the transfer function with respect to the normalized in-plane wavevector $k_x/k_0$ and angular frequency $\omega$. **e-f** Amplitudes (black) and phases (blue) of the transfer function along $\omega = \omega_0$ (white lines in **c-d**) and $k_x = 0$. The solid and dashed lines correspond to the transfer functions of experimental results and the fitting ones with Eq. 3, respectively. The black and blue arrows represent the curves to the left and right *y*-axis.

Figs. 3c-d plot the measured magnitude $|H|$ and phase $\angle H$ of the spatiotemporal transfer function of our sample as a function of the normalized in-plane wavevector $k_x/k_0$ and angular frequency $\omega$, respectively. From Fig. 3c, one can observe that a pronounced resonant mode was excited at $\omega = \omega_0$ where $\omega_0 = 7.09 \times 10^{10}$ rad/s (frequency $f_0 = 11.28$ GHz) for normal incidence. The phase singularity presented in the phase distribution of the transfer function (see Fig. 3d) makes it possible that the amplitude of the transfer function is close to zero when $k_x = 0$ and $\omega = \omega_0$ as shown as Fig. 3c. The discrepancy between the measured and simulated results for the transfer function of the metasurface may be ascribed to the fabrication errors of the sample and the backscattering interference from the environment. Black and blue solid lines in Fig. 3e-f show the amplitudes and phases of the measured transfer function of the sample along $\omega = \omega_0$ and $k_x = 0$, respectively. We see that when $-0.02 \leq k_x/k_0 \leq 0.02$ and $-0.002 \leq \Omega/\omega_0 \leq 0.002$, the amplitudes of the transfer function could be roughly fitted by a linear function with respect to both $|k_x|$ and $|\Omega|$. Additionally, there is ~$\pi$ sharp change for the phase of the transfer function in a narrow region around $k_x = 0$ and $\omega = \omega_0$. This indicates the fabricated sample can be implemented to realize the first-order analogue spatiotemporal differentiation. With $c_x = 15.60 \exp(1.68i)/k_0$ and $c_t = 48.18 \exp(-0.72i)/\omega_0$, the fitting curves of amplitudes and phases with Eq. 3 presented by dashed lines in Fig. 3e-f roughly coincide with those of experimental results. $c_x$ has a phase shift of 2.4 rad from $c_t$, which results in the phase singularity for the transfer function $H(k_x, \Omega)$ and further the zero amplitude with $k_x = 0$ and $\omega = \omega_0$.

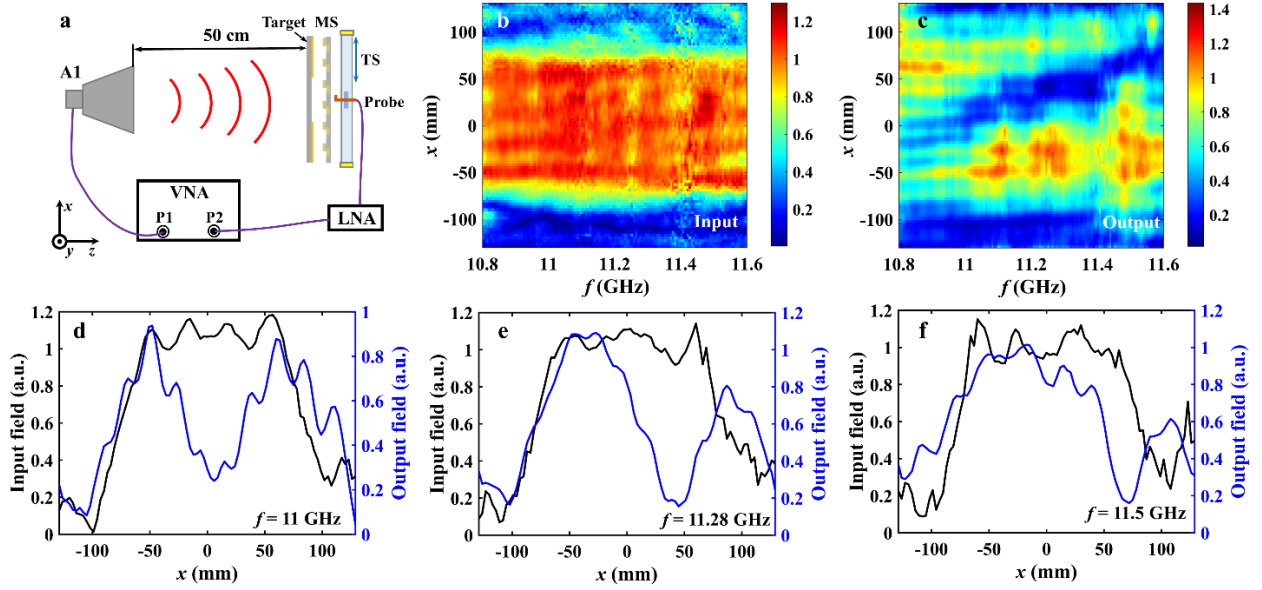

**Fig. 4 Experimental results for the spatial edge detection of the metasurface sample. a** Schematic diagram of experimental setup employed to evaluate the spatial edge detection capability of the metasurface. A1: transmitting antenna, MS: metasurface, TS: motorized translation stage, LNA: low-noise amplifier, VNA: vector network analyzer, P1/P2: ports of VNA. **b-c** Measured amplitude distribution of the input and output field as functions of frequency $f$ and coordinate $x$. **d-f** Measured amplitude profiles of input (black) and output (blue) fields as a function of $x$ at the operating frequency $f$ = 11 GHz, 11.28 GHz, and 11.5 GHz, respectively.

One of the important applications of the first-order analogue spatial differentiation is the edge detection[1], which is widely harnessed in modern automation systems such as Light Detection and Ranging (LiDAR) system, autopilot system, and intelligent medical diagnosis system. The spatial edge detection ability of the metasurface sample was scrutinized with the experimental setup shown in Fig. 4a. A self-made target, which was fabricated by sticking the copper foil tape on two sides of a piece of foam board to form a slit with a width of about 150 mm for the microwave regime, was utilized to test the edge detection capability of the metasurface. The metasurface was placed directly in front of the target. We used the transmitting antenna A1 connected to VNA as the microwave source and displaced it away from the target approximately by a 50 cm distance. A probe connected to a coaxial cable was mounted on a motorized translation stage TS (A036, HWHR) to detect near-field one-dimensional spatial signals just after the metasurface. In the experiments, the motorized translation stage was precisely controlled by a stepping motor controller (ASC-N, HWHR) and moved with a step of 3 mm continuously. The detection space has a width of 270 mm. The low-noise amplifier LNA (TLA-001030G30, THZW) was connected to the probe and VNA to amplify the measured signal from the probe above the noise level. A LabVIEW program was utilized to realize the automatic acquisition of the near-field profiles by controlling the motorized translation stage and processing data from VNA simultaneously. The input (with target in the setup) and output (with target and metasurface both in the setup) field profiles were normalized by the field profile detected by the probe without target and metasurface in the setup, and were displayed in Fig. 4b-c, respectively. Distinguishable slit patterns can be observed in the input field profiles for frequencies of interest shown in Fig. 4b. The slight fluctuations within the slit are attributed to the microwave diffraction. From Fig. 4c, one will find that the edges of the slit are enhanced by different degrees when the operation frequency changes. The first-order spatial differentiation of an ideal edge is δ-function, which will be degenerated to inverted V-shaped pattern for a limited spatial frequency bandwidth. Figs. 4d-f plot the measured amplitude profiles of the input and output fields with respect to $x$ while $f$ = 11 GHz, 11.28 GHz, and 11.5 GHz, respectively. For the resonant frequency $f_0$ = 11.28 GHz, the inverted V-shaped patterns are presented near edges of the slit in the output field profile depicted in Fig. 4e, suggesting the spatial edge detection of the metasurface. When the frequency decreases to 11 GHz, edges of the slit could still be detected. The edge detection function deteriorates with the frequency up to 11.5 GHz due to the nonnegligible deviation of the spatial transfer function from the ideal case.

Then the first-order temporal differentiation function of the metasurface sample is experimentally examined by considering incident pulse envelopes with the Gaussian-like lineshape. The center frequency of the incident temporal pulse is 11.28 GHz and the full width at half maxima (FWHM) of the pulse envelope is denoted by $\Gamma$. Fig. 5 provides experimental measurements of the temporal differentiation for incident pulse envelopes with different $\Gamma$ using the setup in Fig. 4a where the target was removed. Amplitude profiles of output pulse envelopes for the ideal device with $H(k_x = 0, \Omega)$ fitted by Eq. 3 are depicted by blue dashed lines, which exhibit typical M-shaped profiles with zero amplitudes at $t = 0$. The time delay appearing between the output pulse profiles and the ideal ones reduces when $\Gamma$ is increased. For input temporal pulses with $\Gamma$ = 10.3 ns and 16.8 ns, the device experimentally produces an acceptable first-order temporal differentiation.

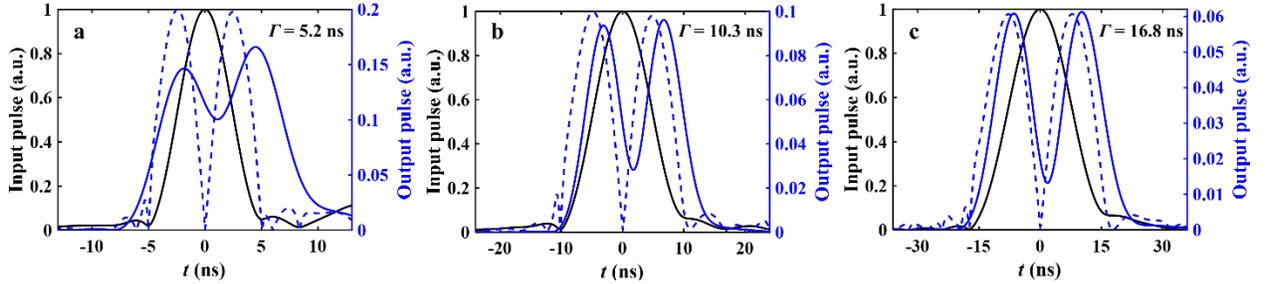

**Fig. 5 Experimental measurements of amplitude profiles of pulse envelopes as a function of time $t$ for Gaussian-like temporal pulses. a** $\Gamma$ = 5.2 ns. **b** $\Gamma$ = 10.3 ns. **c** $\Gamma$ = 16.8 ns. $\Gamma$ is the FWHM of the incident pulse envelope. Black and blue lines represent amplitude profiles of input and output pulse envelopes, respectively. Solid and dashed lines represent the metasurface sample and ideal device with $H(k_x = 0, \Omega)$ fitted by Eq. 3 ($c_t = 48.18 \exp(-0.72i)/\omega_0$), respectively.

**Sharp change detection for spatiotemporal pulses.** Similar to the spatial edge detection function of the spatial differentiator, the spatiotemporal differentiator can be implemented to detect sharp changes of spatiotemporal pulses. With the measured spatial and temporal transfer functions plotted in Figs. 3c and 3d, we can directly inspect the spatiotemporal detection effects of the metasurface sample by considering spatiotemporal pulses with amplitude modulation. Input pulses with spatial widths of 600, 300, 150, 120, and 60 mm are given in Fig. 6a, at an identical temporal width 20 ns. Fig. 6d shows the amplitude distribution of the output spatiotemporal pulse corresponding to Fig. 6a, suggesting that the metasurface could detect sharp changes in the spatiotemporal pulses. When the spatial width of the pulse decreases, the edges in the spatial domain become blurred. The spatial edges of the pulse with the spatial width 60 mm are difficult to be distinguished, indicating the spatial detection resolution of the metasurface sample is about 120 mm. Then we consider input pulses with different temporal widths of 20, 10, 4, 2, and 1 ns shown as Fig. 6b, where the spatial widths of these pulses are fixed at 1000 mm. The amplitude distribution of the corresponding output pulse is provided in Fig. 6e. Similarly, the edges in the temporal domain become blurred when the temporal width of the pulse is reduced. The temporal edges of the pulse with a temporal width of 1 ns cannot be recognized. Therefore, the temporal detection resolution of our device is estimated at about 2 ns. The spatiotemporal differentiator would also detect sharp changes in complex spatiotemporal pulses such as the badge of Zhejiang University shown in Fig. 6c. Fig. 6f depicts the amplitude distribution of the corresponding output pulse. One can identify sharp changes in the spatiotemporal pulse.

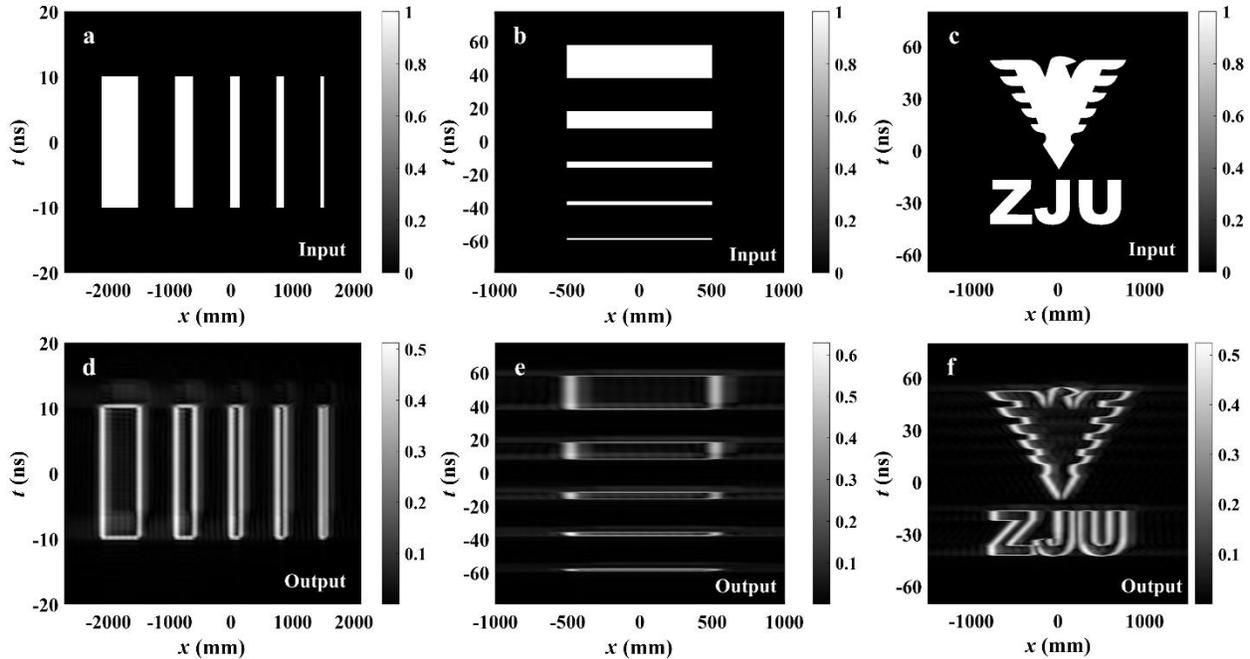

**Fig. 6 Theoretical analysis of sharp change detection for spatiotemporal pulses. a** Input rectangular spatiotemporal pulses with spatial widths of 600, 300, 150, 120, and 60 mm, from left to right. These pulses have the same temporal width 20 ns. **b** Input rectangular spatiotemporal pulses with temporal widths of 20, 10, 4, 2, and 1 ns, from top to bottom. These pulses have the same spatial width 1000 mm. **c** Input spatiotemporal pulse as the badge of Zhejiang University. **d-f** Output spatiotemporal pulses corresponding to **a-c**, respectively.

**Discussions**

In summary, the analogue electromagnetic spatiotemporal differentiator based on the metasurface was first demonstrated experimentally. The spatiotemporal transfer function of the device working at microwave frequencies mimics that of the ideal first-order spatiotemporal differentiation by the unidirectional excitation of SSPPs in the bilayer metal grating, which is endorsed by simulations and experiments. We verified that the differentiator fabricated could perform spatial edge detection to a slit with a width of about 150 mm in the vicinity of the working frequency 11.28 GHz. Experimental measurements of temporal processing results of the device indicate that the device would realize high-accuracy first-order temporal differentiation for input Gaussian-like pulse envelopes with FWHM over 10 ns. The spatiotemporal differentiator demonstrated in this work could detect sharp changes in spatiotemporal pulses. The spatial and temporal detection resolutions of the device were estimated as 120 mm and 2 ns, respectively. We further confirmed the sharp change detection capability of the differentiator for complicated spatiotemporal pulses by considering the badge of Zhejiang University as the input pulsed field. The spatiotemporal differentiator implemented in this work would be conveniently replicated in terahertz, infrared, and visible regimes by leveraging the excitation of SSSPs or SSPs.

Recently, researchers have exploited spatiotemporal differentiators based on metasurfaces[57–59] and multilayered structures[60], which are much more compact than traditional pulse shapers[61,63,64] with indispensable spatiotemporal Fourier transfer components, to generate spatiotemporal optical vortices (STOVs) with transverse OAM. In Supplementary Note 1, we show that our first-order spatiotemporal differentiator can not only give rise to the spatial and temporal profiles of a single rectangular-shape pulse input but also stamps a transverse OAM with a fractal topology charge number ($l = 0.58$) for the output pulse. This is equivalent to add additional information channel for the output field when compared with a high-order spatiotemporal differentiator. It is worthy deeper investigating in the future on OAM-carried analogic optical spatiotemporal computing by configuring the transfer functions in the four-dimensional space as STOVs with transverse OAM have been widely applied in the fields of optical communication, super-resolution imaging, quantum key distribution and so on[58]. Benefiting from its advantages including extraordinary integration capability, low-energy consumption, and natural parallel processing ability, the analogue electromagnetic spatiotemporal differentiator demonstrated in this work is believed able to find kaleidoscopic applications in Internet of Things (IOT), electromagnetic imaging, quantum communication, to name a few.

**Acknowledgements**

The authors are grateful to the partial supports from the NSFC of Zhejiang Province (LXZ22F050001), the NSFC (62075196, 61775195 and 61875174), the National Key Research and Development Program of China (No. 2017YFA0205700) and the Fundamental Research Funds for the Central Universities.

**Author Contributions**

Y.Z. conducted the most of the experiment and theoretical work. J.Z. and Z.X. conducted part of the experiment. Y.S., Y.W., Y.D., and S.Z. carried out part work of the data analysis. Y.M. supervised the work. All the authors participated in the writing of the manuscript.

**Competing interests**

The authors declare no competing interests.